\documentstyle[twocolumn,aps]{revtex}
\begin{document}

\title{Optical Bistability in Colloidal Crystals}

\author{E.~Lidorikis, Qiming Li, and C.~M.~Soukoulis}

\address{Ames Laboratory and Department of Physics and Astronomy,\\
Iowa State University, Ames, IA 50011\\}
\author{\parbox[t]{5.5in}{\small
We present a  one dimensional model for the nonlinear response of 
a colloidal crystal
to intense light illumination along a high symmetry direction. The strong coupling between light and the 
colloidal lattice,  via the electric gradient force acting 
upon the particles,  
induces a novel large optical nonlinearity. We obtain bistable behavior when 
the incident frequency is inside the stopband of the periodic structure,
with decreasing switching intensity as the frequency increases. The
transmission characteristics and the magnitude of the switching 
threshold intensity are also in good agreement with 
a recent experiment.\\
\\
PACS numbers: 42.65.Pc \quad 82.70.Dd \quad  42.70.Qs  \quad 78.66.-w }}
\maketitle
%\newpage

\section{Introduction}

Photonic band gap (PBG) materials \cite{bib1} do not allow propagation of
electromagnetic waves within a certain frequency range, thereby opening the
possibility of studying new physics within the gap. In addition, many
novel applications of these PBG crystals have been proposed, with operating
frequencies ranging from microwaves to the optical regime\cite{bib1,bib2}.  
Structures exhibiting full photonic band gaps in the microwave\cite{bib3}, 
millimeter\cite{bib4} and
submillimeter\cite{bib5} regimes have already been fabricated, but scaling these
structures down to the optical regime has remained a challenge.  One way to 
construct PBG crystals in the optical regime is by growing polystyrene
colloidal
crystals\cite{bib6,bib7,bib8,bib9}, which have lattice spacing comparable to the
wavelength of light.  Such colloidal crystals do not exhibit a complete 
PBG, because the concentration and the index of refraction of the 
polystyrene spheres relative to water are not yet sufficiently high.
However, they are very useful in studying PBG effects seen only 
in particular directions. In addition, they can be used in nonlinear   
optical studies.  It is expected that the PBG effects can be strongly
affected by nonlinear optical effects.  
In particular, it has been shown that intensity-dependent
index of refraction can cause a shift in the locations of the band
gap\cite{bib10}.  That is, if
the index of refraction of either the suspended polystyrene spheres 
or water is intensity-dependent, then the width and the position of the 
stopband (or gap) will change under intense illumination.  
For example, a decrease of the index of refraction in water upon illumination
will widen the gap, and, therefore, inhibits the
propagation of the probe
beam.  Such an optical switching for light control is of great interest to 
the optics community.

In a recent experiment\cite{bib7}, optical switching and optical bistability
were observed, when intense light was transmitted through a
colloidal crystal.  Simple switching was observed near the low-frequency end
of the stopband, whereas bistability and multistability occurred at 
the center or near the high frequency end. The switching
threshold was found to decrease as the incident frequency increased. These 
aspects are inconsistent with the response of a material with the conventional 
intensity-dependent nonlinearity\cite{bib7}. The measured nonlinear coefficient 
n$^{^{\prime}}$ inside the transmission band, $4 \times 10^{-10} cm^2/W$, 
is also several orders of magnitude larger than the electronic 
nonlinearity of both materials. 

In this paper, we present a one dimensional model for the nonlinear response
of a colloidal crystal to intense light incident along a high 
symmetry direction, based on the electrostriction mechanism\cite{bib7}.
Light is strongly
scattered by the periodic arrangement of the colloidal particles 
inside the crystal, thus creating a spatially varying field.  
The polystyrene spheres, polarized by the electric field,
will move in response to the electric gradient force. Such a structure
change in turn will alter the propagation of light. The optical  
response is thus controlled by the the stationary configuration that results 
from the balance between the elastic and the electric gradient forces.
We assume that the electric gradient forces are not strong enough to
destroy the polystyrene spheres' crystalline structure and furthermore, 
that the
changes in interparticle separation are small compared to the mean
interparticle separation. Assuming the incident wave can be approximated
as plane waves due to the large beam spot size, the structural change 
induced by light incident along a high symmetry direction will be primarily
in the propagating direction. We neglect possible transverse effects 
and describe the three dimensional lattice by a one
dimensional harmonic lattice model.
Numerical calculations of the transmission characteristics, based on 
known physical properties of the colloidal crystal and a 
simplification to a layered structure, show good agreement with 
experiment. In particular, we find bistable behavior inside the 
stopband at intensities comparable to the observed switching threshold.
The switching threshold is found to decrease as the incident 
frequency increases.  We need to emphasize that such a 
nonlinearity necessarily depends on the exact stationary configuration of the 
lattice and therefore cannot be described with a simple effective 
intensity- and/or frequency-dependent dielectric constant.

This paper is organized as follows. In section II, we introduce
our one-dimensional lattice model for the optical nonlinearity 
in colloidal crystal.  In section III,  we present results of calculations 
on the optical bistability and compare them with experiments.  
Conclusions and discussions are presented in section IV. 

\section{One Dimensional Model of Optical Nonlinearity}

In general, wave propagation in periodic structures is a complex phenomenon.
Three dimensional scattering of light plays an important role in determining
the nonlinear optical response of a colloidal crystal to incident light. 
However, simplification is possible if,  a) the incident wave is 
plane-wave-like; b) the light is normally incident upon a high symmetry plane
of the crystal lattice, and c) no transverse instability exits. 
Under these conditions, the structure can be viewed as a layered system. 
One dimensional modeling of the optical response is expected
to be appropriate with correctly calculated physical parameters. 
The first condition ensures all spheres within one layer are equivalent, hence
there should be no lateral lattice displacement, as required by symmetry.
The second condition makes the layered structure more distinct since the
distance between the layers is large. The third condition essentially
requires that the structure is stable under illumination. 

The colloidal crystal used in experiment\cite{bib7} had a face-centered-cubic structure 
formed by polystyrene spheres ($n_1$=1.59) of approximately $d_1$=120 nm 
in diameter, at concentration f around 7\%, dispersed in water ($n_2$=1.33).
Light was normally incident upon the [111] plane which was parallel to the 
surfaces of the container.  Due to the relatively large spot size, we  
approximate the incident wave as plane-waves. To a low intensity incident wave
from the [111] direction, the fcc colloidal crystal acts essentially as a 
Bragg reflector (linear regime). The system is naturally simplified as a one 
dimensional bilayer structure consistent of alternating segments 
of polystyrene sphere layers ( a mixture of polystyrene spheres and water 
with total thickness d$_1$)
and pure water layers.  With sphere concentration of 6.9\%, the distance between 
the polystyrene sphere layers\cite{bib11} is R$_0$=216 nm. The thickness of the
water layer is then d$_2$=R$_0$-d$_1$=96 nm. The average index of refraction 
of the polystyrene layer is estimated\cite{bib12} to be 1.36.  Since the  
sample thickness is L=100 $\mu$m, the total number of bilayer units is
N=L/R$_0$=463. The linear (zero intensity limit) transmission coefficient 
versus the wavelength in the [111] direction is shown in Fig.~1, 
calculated with the parameters mentioned above. 
Notice that there is excellent agreement\cite{absorp} between our  
theoretical results of Fig.~1 and the experimental results of Fig.~2 in
Ref. 7. This shows our model parameters describe very well the linear 
transmission of the colloidal crystal.  

To illustrate that the nonlinear response of the colloidal crystal 
cannot be described by a one dimensional layered model with the conventional
Kerr type nonlinearity (intensity-dependent  
dielectric constant), we show in Fig.~2 the nonlinear response of such a  
system, assuming that the effective index of refraction n$_2$ of 
the ``water" has the form,  $n_2 = n_2^o + n^{\prime} \vert E \vert^2$.
n$_2^o$=1.33 is the linear index of refraction of the medium and
the nonlinear coefficient is taken to be the experimentally measured 
nonlinearity, n$^{\prime} = 4 \times 10^{-10} cm^2/W$. Taking the
propagation direction to be the $z$ direction,  
we can solve the propagation of light governed by the following wave equation
\begin{equation} 
\frac{d^2 E}{dz^2} + \frac{n^2 (z) \omega^2}{c^2} E = 0   ,
\end{equation}
where n(z) is the index of refraction of our model which consists of alternating
layers of linear and nonlinear medium, as specified above.  
Indeed, bistable behavior is obtained, as can be clearly seen in Fig.~2.
The existence of such bistability phenomena in distributed feedback structures 
with intensity-dependent dielectric constants were predicted\cite{winful}
theoretically and their properties have been investigated 
intensively\cite{desterke}.  
Similar bistable behavior has been seen\cite{bib13,bib14,bib15} in the
discrete case of the electronic version  of Eq.~(2).  However, 
the threshold intensity for the onset of the bistable behavior is in the order
of 10 MW/cm$^2$, at least three orders of magnitude larger 
than the experimentally measured value of about 5 kW/cm$^2$.  
We see that a simple solution of Eq.~(1) with an effective 
intensity-dependent nonlinearity indeed produces 
bistable behavior,  but its predictions for the incident intensity
threshold are unrealistically high. Clearly, a novel form of nonlinearity 
must be in action. 

An interesting mechanism due to electrostriction was proposed\cite{bib7} to be 
responsible for the nonlinear behavior in the colloidal crystal. 
In the absence of 
light illumination, the short-range screened electrostatic repulsive
forces\cite{bib16} between the spheres balances the weak long-range
attractive force of Van der Waals type and produces an equilibrium
configuration for the polystyrene spheres, with nearest neighbor separation
$S_0=\alpha/\sqrt{2}$ where $\alpha$ is the fcc lattice constant.
When intense electromagnetic (EM) wave is incident upon the crystal, the 
dielectric spheres get polarized by the electric field and move in
response to the gradient force \cite{bib16} from the spatially varying field.
This in turn alters the propagation of EM wave. Consequently, optical 
nonlinearity result. Ultimately, the optical response under a given illumination
is controlled by the steady state lattice configuration which has to be
determined by the balance of the elastic and electric 
gradient forces. 

To determine whether the electrostriction mechanism 
is indeed responsible for the experimental observations, we have to  
solve simultaneously wave propagation equations and lattice 
dynamics incorporating both elastic and electric gradient forces.
Here we propose a simple one dimensional lattice model
that can be solved straightforwardly but still contains the
essential physics to account for the optical bistability 
observed in experiment. We have argued that under the 
experimental condition, a one dimensional layered model is 
appropriate to describe the transmission of wave along the
propagation direction. The linear transmission property of
this one dimensional lattice, consisting of alternating layers
of polystyrene and water, has already been described (Fig.~1). 
To model the nonlinear response, we need knowledge of the lattice
dynamics which is governed by the elastic and electric gradient 
forces. 

We assume that for small fluctuations around the equilibrium 
configuration,  the harmonic approximation is correct. We can then describe 
the motion of polystyrene spheres as if they were connected with each other 
with ideal springs. The force constant k of the springs can be roughly
estimated by linearizing the screened electrostatic repulsive 
force\cite{bib19},
\begin{equation}  
F_{el}=\frac {(Ze)^2} {4\pi \epsilon R^2}
\frac {1+\kappa R} {1+\kappa a} \exp[-\kappa (R-a)],
\end{equation}
 at the layer
equilibrium position $R=R_0$. This leads to an expression 
$F_{el}=F_{harm}=k \Delta R$, where $\Delta R = R-R_0$  is 
the displacement from the equilibrium position.
Z is the 
number of charges on the particle, $\kappa$ is the inverse screening length, and
$a$ is the radius of the particle. Using $\epsilon=1.33\epsilon_0$,  
$a$ = 60 nm, $R_0$ = 216 nm, and assuming typical values\cite{bib16} of 
Z=1000e$^-$ and $\kappa =5\times 10^7 m^{-1}$, we obtain k = 1.8$\times 10^{-4} $N/m.
This corresponds to a bulk moduli B$\sim$ k/$R_0 \sim$1000 N/M$^2$, 
a reasonable 
value for colloidal crystals\cite{bib16}.  As we will see
later, the electrostriction nonlinearity is inversely proportional to 
k. Only the order of the magnitude of k is relevant for 
our purpose. For definiteness, we take k = 1.8$\times 10^{-4} $N/m 
in the following calculations. 

The  gradient force on a sphere F$_{gr}$ along the propagation
direction $z$, can be calculated by taking the spatial 
derivative of it's  polarization energy, i.~e., 
$F_{gr}=-d(U_p)/dz$. A crude estimate of this force is 
\begin{equation}
F_{gr}\simeq 4 \pi n_1^2 \epsilon_0 \frac{m^2-1}{m^2+2} a^3
\frac{1}{2}\frac{\Delta \vert E \vert ^2}{2a}
\end{equation}
where $m^2=(n_1/n_2)^2$ is the dielectric contrast between polystyrene
spheres and water, $a$ is the  radius of the spheres,  and the factor
$\frac{1}{2}$ comes from averaging over a time period. $\Delta \vert E
\vert ^2$ is the field intensity difference across a sphere's diameter.   
For our model's parameters this gives $F_{gr}\simeq C \Delta \vert E
\vert ^2$, $C=2.2\times 10^{-26} Nm^2/V^2$.

The optical response of the colloidal crystal is determined by
the steady state configuration. In our one dimensional model, this is 
reflected as the steady state lattice configuration representing the 
configuration of the layers. Taking nearest neighbor interactions 
only and denoting by $\Delta R_n$ the change from the
equilibrium separation of particles n and n+1, we have for the 
steady state that
\begin{equation}
F_{gr} = -F_{harm} =- k (\Delta R_n - \Delta R_{n-1}).
\end{equation}
The gradient force $F_{gr}$ on each polystyrene layer has to be calculated 
from the electric field distribution via Eq.~(3)  for the given lattice 
configuration $\{R_n\}$.

The transmission characteristics are obtained by solving Eqs.~(1) and (4)
self-consistently through iteration, for a given input. In actual calculations,
however, this is done for a given output because the presence of bistable 
or multistable behavior.  In a nonlinear one-dimensional model, each 
output corresponds
to exactly one solution,  while a given input may correspond to more than one
output solutions (bistability). The input intensity can be 
reconstructed once the transmission coefficient is calculated after solving 
Eq. (1). We start with the equilibrium configuration in the absence of light
in which all the layers 
is equally spaced with distance $R_0$. The wave field E(z) is then calculated 
from Eq. (1),  with n(z) given by the exact one-dimensional lattice 
configuration $\{R_n\}$. n(z) equals to 1.36 if z is  
in the polystyrene layer and 1.33 otherwise. The gradient force and the 
elastic force is then calculated and $R_n$ is increased or decreased depending
the direction of the total force. The wave field is then recalculated and
accordingly $\{R_n\}$ readjusted. This iteration procedure continues until
the total force vanishes on each polystyrene layer. The final configuration
will be the steady state configuration,  and the corresponding field represents
the actual optical response of the system. Twenty iterations are usually required before a steady state self-consistent
configuration is achieved. 

We point out that the
present situation is analogous to the problem of an electron moving  
in a one-dimensional harmonic lattice with electron-lattice interactions.
Such an analogy may help to understand the nonlinear optical response 
when the frequency is inside the stopband. We comment that neither in
the 
polystyrene spheres nor in the water have we 
assumed any intrinsic nonlinearity. The nonlinear response of the colloidal
crystal is entirely due to the coupling between the light and the lattice. 
In principle, such coupling exists in all materials. But the extreme 
softness of colloidal lattices relative to conventional crystals, reflected 
in the small value of the effective spring constant k, makes 
the observation of nonlinear effects  possible in
these materials. 

\section{Optical Bistability}

As a first check of our model, we have numerically calculated the sign and the
strength of the effective nonlinearity for a frequency ($\lambda$=514 nm) 
inside the transmission band.  We found that the colloidal crystal linearly
expands with the incident intensity of the EM wave,
with a slope of about $1 nm$ per $30 kW/cm^2$. This corresponds to a
relative linear expansion of the order $\Delta L/L \simeq 10^-5$, which
is quite small as required by our harmonic assumption. The resulted   
phase shift in the transmitted wave can be related  to  
an effective positive nonlinear index of refraction by
\begin{equation}
\frac{\omega}{c}(n_2-1)\Delta L\simeq\frac{\omega}{c} 
n^{\prime}\vert E_0 \vert^2 L
\end{equation}
where $\vert E_0\vert$ is the incident intensity. For our system we
estimate 
n$^{\prime} \simeq 10^{-10}
cm^2$/W. This is in excellent agreement with the experimental value\cite{bib7} 
of $n^{\prime} \simeq 4 \times 10^{-10} cm^2$/W,  considering 
that the value of the force constant is only estimated with typical 
values of physical properties for colloidal crystals. We find that within the
transmission band, the nonlinearity  n$^\prime$ scales   
almost linearly with 1/k, but with no appreciable frequency
dependence. The experimentally observed nonlinearity can be matched 
with the choice k$\simeq 4.4\times 10^{-5} N/m$ and a corresponding
bulk moduli B$\sim 250 N/m^2$. For definiteness we continue to use the 
initial estimated value of k. Changing value of k amounts to 
rescale the light intensities, since the actual contraction or expansion
is controlled by the ratio of the elastic force and the gradient force, 
ie, only the ratio of the k and light intensity matters. 

Multistability and switching threshold intensities are also 
correctly predicted within our model for frequencies inside the band
gap.   The local expansions and
contractions of the lattice under illumination are the origin of the 
bistable behavior.  Normally, transmission is forbidden in the gap 
of a periodic system. However, 
lattice distortion allows the existence of 
localized modes in the gap.  Under appropriate conditions, the coupling of 
these localized modes with the radiation can produce resonant transmission. 
This is clearly seen in Fig.~3, where the local 
lattice expansion (a), the field intensity averaged in each sphere (b), and
the intensity gradient (c), are shown as 
a function of the lattice plane, exactly at a transmission resonance 
for a frequency inside the stopband.  
Notice that there is a strong lattice deformation (solid
curve in Fig.~3a) at the middle of the crystal, sustained by the strong field
intensities (solid curve in Fig.~3b) and  the intensity gradient.
Similar behavior is seen for the case of the second transmission resonance 
(dotted lines in Fig.~3).  This work clearly shows that there exists 
a strong light-lattice interaction, giving  rise to lattice
deformations which in turn produces localized solutions as ``soliton-like"
objects\cite{bib17,bib18}.
When these ``soliton-like" objects appear symmetrically in the crystal, a
transmission resonance is expected.  Also, the longer the wavelength and the
higher the multistability order are, the larger the maximum values of these
deformations become.  The bistable behavior originates from these 
field-distribution-specific structure 
changes. We point out that the total expansion of the lattice is still 
relatively
small, generally in the order of 80 to 200 nm for each ''soliton-like" 
object present in  the structure.  

The transmission characteristics are shown in Fig.~4,
for four different wavelengths as were indicated in Fig.~1.  We see that
our model captures the most essential  features of the nonlinear
response of the colloidal crystal, as  
compared with the experimental results presented in Fig.~3 of Ref.~7.
Notice that this model correctly predicts the magnitude of the switching 
threshold intensities, they are of
the order of (20-40) kW/cm$^2$ and not of the order of 10$^4$ kW/cm$^2$ that
the simple model with an intensity-dependent dielectric 
constant predicts (see  Fig.~2) .  The switching threshold
intensities get smaller as we move from the long to the short wavelength
side of the stopband, in agreement with experiment\cite{bib7}. 
Bistability is observed when the lattice is distorted enough so to sustain
a localized mode. This will happen if the local expansion is large
enough to locally
shift the effective gap to longer wavelengths \cite{bib19}. 
Thus, the closer the incident frequency is to the small wavelength side
of the gap, the smaller the lattice distortion needed to onset
bistability, and thus the smaller the switching powers are.

Discrepancy with the experimental data is found for large incident intensities
and  in the low frequency side of the gap.  Multistability was
observed only in the high frequency side while for midgap frequencies
the crystal is bistable and for low frequencies it is non bistable
\cite{bib7}. Also, at high intensities  only instabilities 
were observed experimentally, in contrast to our model that predicts
 multistable
behavior for all gap frequencies and all intensities. 
However, it is for the long wavelengths and the high intensities that
the required lattice expansions get unrealistically large. The total
lattice expansion versus the transmitted intensity are shown in 
Fig.~5, for the four wavelengths indicated in Fig.~1. Every local maximum  
in these curves corresponds to a transmission resonance.
Since the
crystal can not expand more than a certain maximum limit, an external
pressure must be inserted into our model to limit its expansion.
Numerical studies incorporating an external pressure show that while
multistability is still predicted for all frequencies, the required
local expansions and contractions, (with total expansion being constant
and limited), and light intensities are much larger, making the starting
assumption of a slightly perturbed harmonic lattice invalid. With large
lattice distortions, approximation to a one dimensional structure also
becomes questionable, and this may be the main reason for the discrepancy. 
The neglect of light absorption in water may also be a contributed 
factor to the discrepancy. Absorption reduces the light intensity nonuniformly,
and thus may affect the nonlinear response. 

\section{Discussions and Conclusions}

We have shown that several essential features of the nonlinear response
in colloidal crystals can be accounted for by a simple one dimensional model 
that incorporates the lattice distortions under intense light illumination.
In this one dimensional model, the colloidal crystal is simplified as 
a one-dimensional layered system consisting of alternating layers of 
polystyrene spheres and water. The polystyrene layers represent high symmetry 
planes of the colloidal crystal normal to the propagating direction and are
modeled as elastic media deforming under the act of the gradient force from
the electric field. Based on physical properties of the 
colloidal crystal, we are able to estimate the effective elastic spring 
constant. The wave equation of the electric field and the lattice dynamics
of this one dimensional systems is then solved simultaneously to obtain 
the steady state response. We are able to obtain the correct order of  
 magnitude of the effective nonlinearity within the transmission
band and the switching intensity for optical bistability within the 
stopband.  The trend that this switching intensity decreases as the frequency
increases across the stopband is also reproduced. Although it seems  
surprising that a simple one-dimensional model works when three-dimensional
scattering of light plays an important rule, detail considerations suggest
this simplification should be appropriate under the experimental condition.

In conclusion, we have established with a simple one dimensional model
that the light-lattice coupling via 
electric gradient force underlies the large optical nonlinearity 
observed recently in colloidal crystals. Such a coupling alone can
produce bistability and multistability with switching
threshold intensities and transmission characteristics in good agreement 
with experiment. Given the unique large nonlinear response,  colloidal crystals
may prove to be very useful for future studies of nonlinear effects 
in PBG materials in the optical regime. 

\begin{acknowledgments}

Ames Laboratory is operated for the U.S. Department of Energy by Iowa
State University under Contract No. W-7405-Eng-82. This work was supported by 
the director for Energy Research, Office of
Basic Energy Sciences, and  NATO Grant No. CRG 940647.
\end{acknowledgments}

\begin{figure}
\caption{Transmission coefficient versus the wavelength in
the [111] direction
of a colloidal crystal.  The width of the polystyrene sphere layer is d$_1$=120
nm, with an
effective [12] index of refraction $\bar{n}_1 \simeq$1.36, while the width
of the water layer
is d$_2$=96 nm with n$_2$=n$_2^o$=1.33.  The number of bilayer units is N=463,
which gives a sample 
thickness L=100$\mu$m.  All the parameters used are in agreement with
experiment.  The arrows
indicate the wavelengths of the incident light at which the transmission
plots in Fig.~4 were
obtained.}
\end{figure}

\begin{figure}
\caption{Transmitted intensity versus incident intensity for
a simple nonlinear bilayer system with n$^{\prime} = 4\times10^{-10}$cm$^2$/W
(in water) for two values of the
incident wavelength.  This model gives unrealistically high values for the
threshold
incident intensities.}
\end{figure}

\begin{figure}\caption{Local lattice expansion (a), Field
intensity averaged in each sphere (b), and Field intensity gradient (c),
 as a function of the lattice plane for $\lambda$=579 nm.
 Solid and dashed curves correspond to the first and second
transmission resonances.}
\end{figure}

\begin{figure}\caption{Transmitted intensity versus incident
intensity for four different wavelengths as were indicated in
Fig.~1.}
\end{figure}

\begin{figure}\caption{Total lattice expansion versus transmitted
intensity for the four wavelengths indicated in Fig.~1. No external
pressure is assumed.}
\end{figure}

\begin{references}
\bibitem{bib11}  For an fcc lattice of spheres of filling ratio f, with diameter
d$_1$ and lattice constant $\alpha$, the following relation is obeyed,
f=$\protect 4 \frac{(4\pi
/3)(d_1 /2)^3}{\alpha ^3}$.  The distance between the [111] planes in an fcc
lattice is
R$_0 = \alpha /\sqrt{3}$, and since d$_1$=120 nm, one easily gets that
R$_0$=88.64/f$^{1/3}$ nm.  So for f=6.9\%, R$_0 \simeq$216 nm.
\bibitem{bib12}  
The polystyrene layer is a mixture of polystyrene spheres with diameter d$_1$
and water. To better approximate the experimental system, it is
appropriate to use an average index of refraction  $\bar{n}_1$. 
The volume fraction f$_1$ of the spheres within the polystyrene 
layer is f$_1$=R$_0$f/d$_1$ = 12.4\%.  From 
$\bar{n}_1 = n_1 f_1 + n_2 (1-f_1 )$, we determine $\bar{n}_1 \simeq$1.36.
\bibitem{absorp}
The attenuated transmission in Figs.~2 and 3 in Ref.~7 is due to absorptions in
water which we neglect in this study.
\end{references}
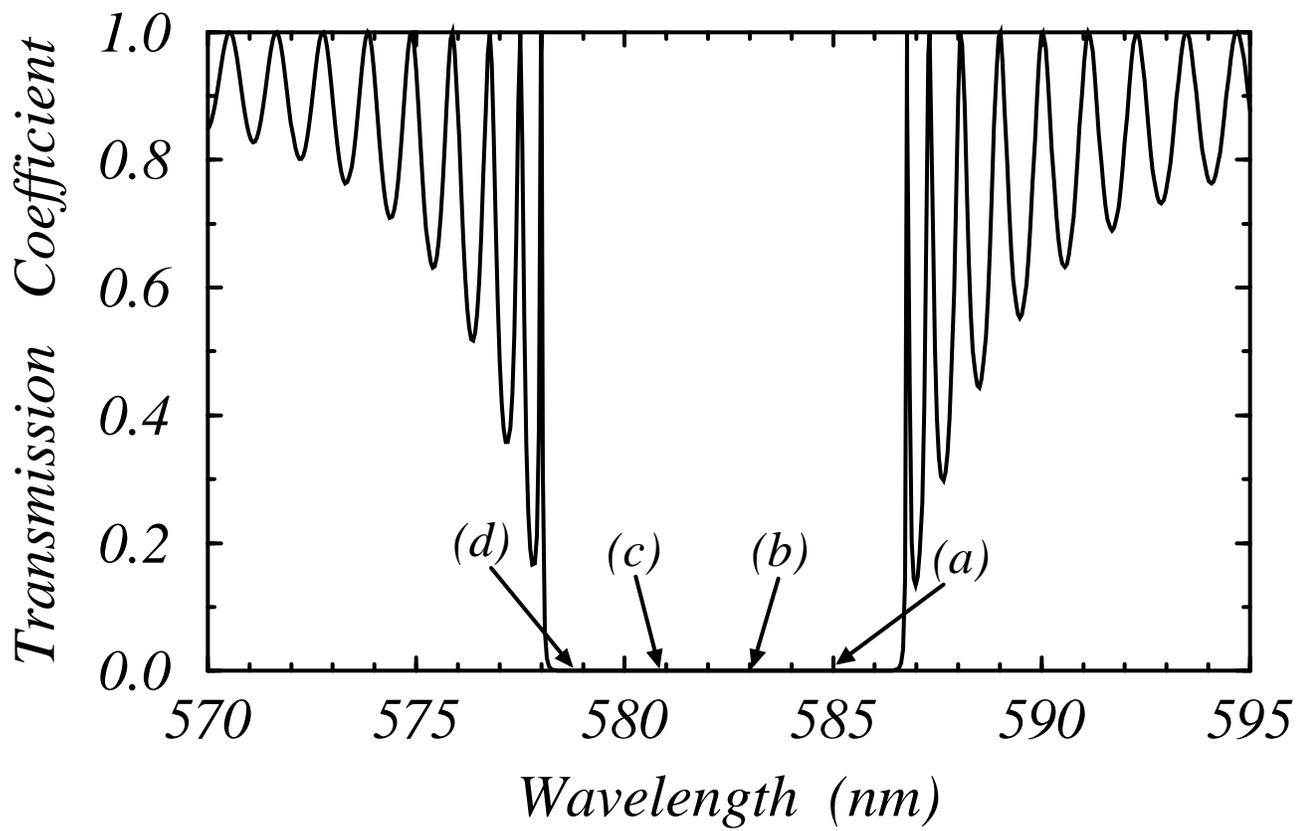
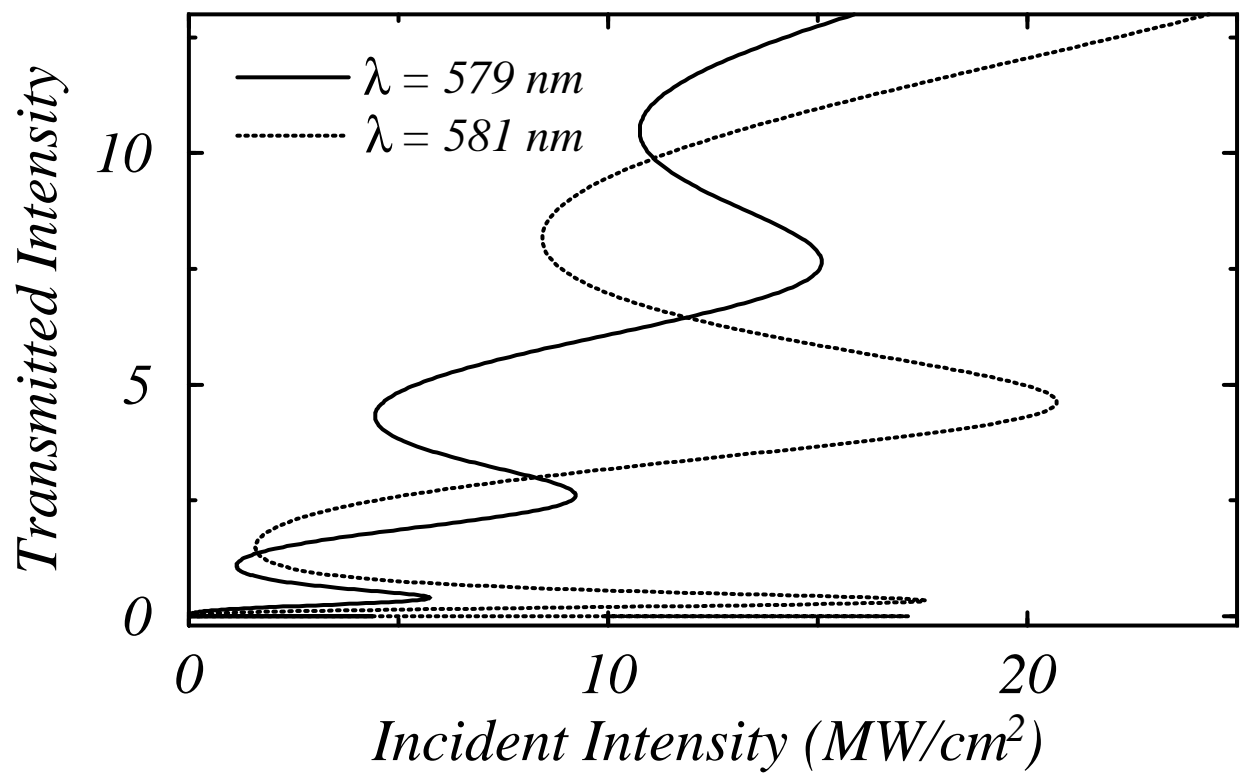
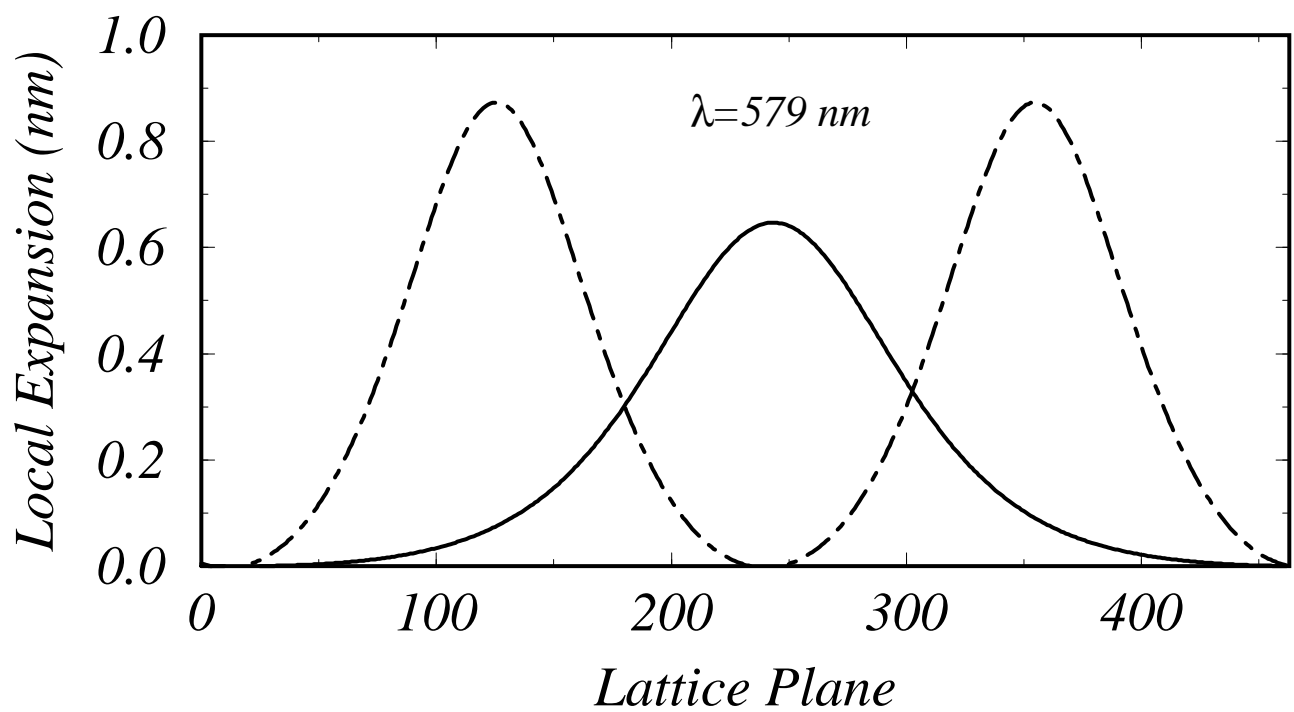
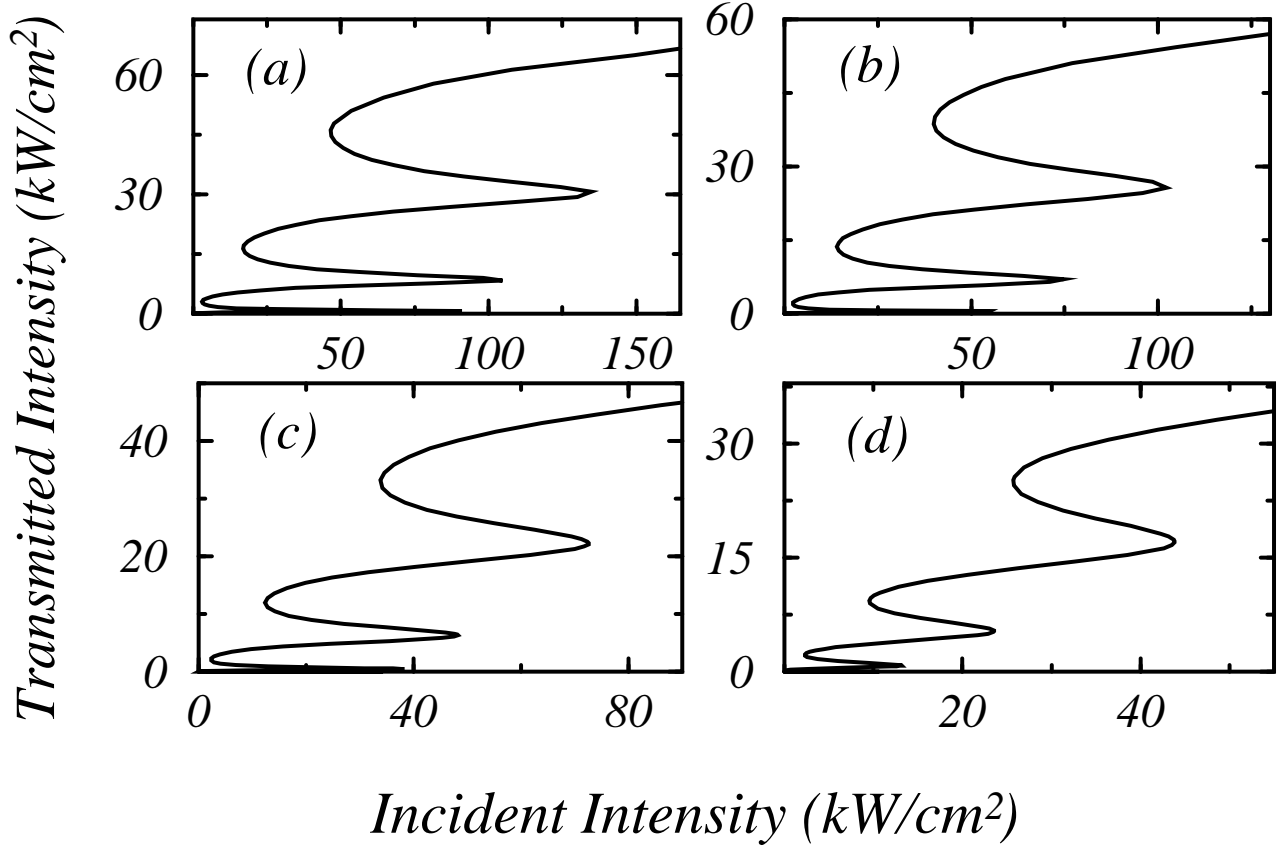
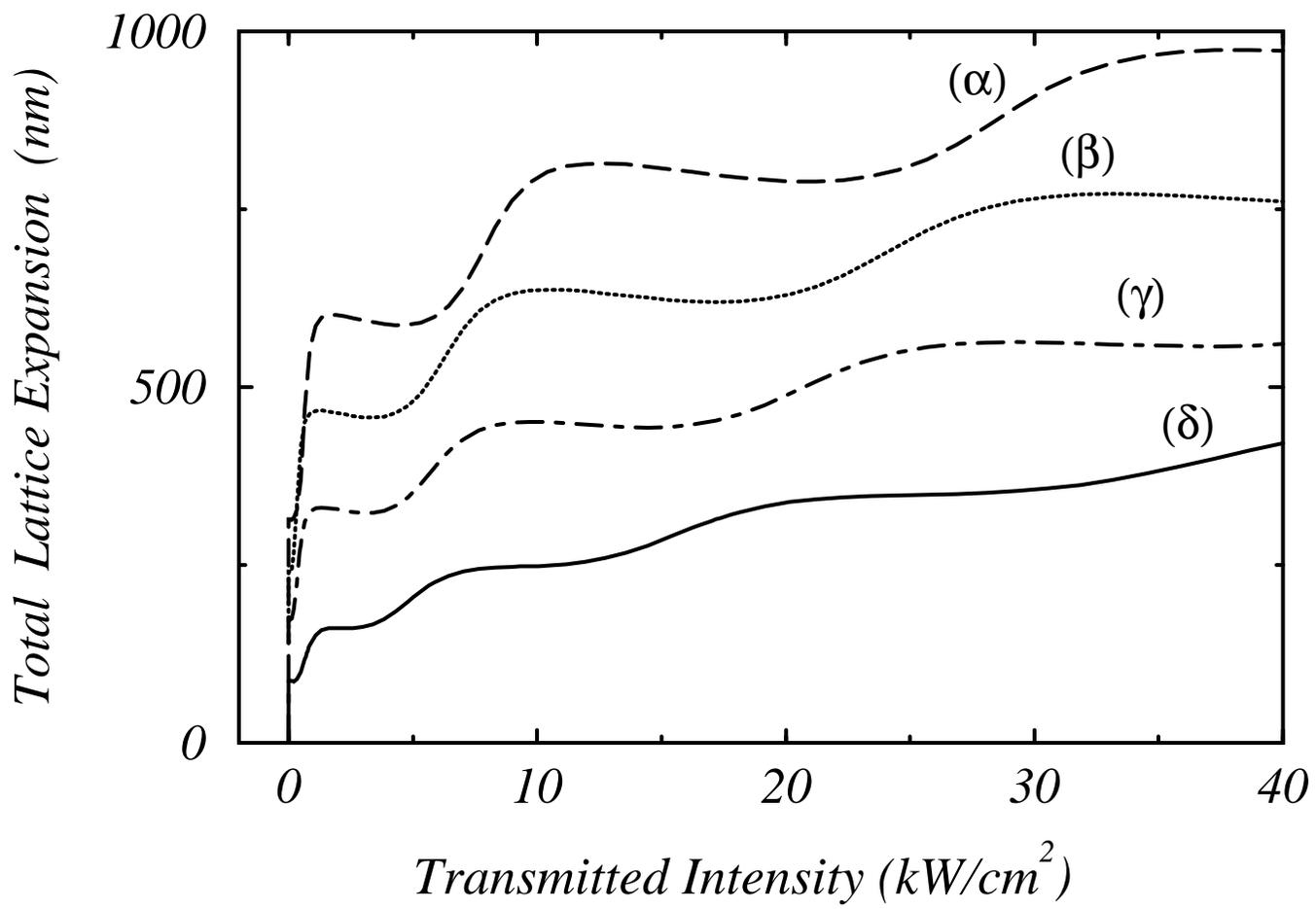
\end{document}